\documentclass[journal = ancac3,manuscript=letter]{achemso}

\usepackage{graphicx}
\usepackage{float}
\usepackage{color}

\author{Jack Hellerstedt}
\altaffiliation{These authors contributed equally to this work}
\affiliation{Institute of Physics of the Czech Academy of Sciences, v.v.i., Cukrovarnick\'{a} 10, 162 00 Praha 6, Czech Republic}
\alsoaffiliation{Regional Centre of Advanced Technologies and Materials, Palack\'{y} University, \v{S}lechtitel\^{u} 27, 78371 Olomouc, Czech Republic}
\author{Ale\v{s} Cahl\'{i}k}
\altaffiliation{These authors contributed equally to this work}
\affiliation{Institute of Physics of the Czech Academy of Sciences, v.v.i., Cukrovarnick\'{a} 10, 162 00 Praha 6, Czech Republic}
\alsoaffiliation{Regional Centre of Advanced Technologies and Materials, Palack\'{y} University, \v{S}lechtitel\^{u} 27, 78371 Olomouc, Czech Republic}
\author{Martin \v{S}vec}
\affiliation{Institute of Physics of the Czech Academy of Sciences, v.v.i., Cukrovarnick\'{a} 10, 162 00 Praha 6, Czech Republic}
\alsoaffiliation{Regional Centre of Advanced Technologies and Materials, Palack\'{y} University, \v{S}lechtitel\^{u} 27, 78371 Olomouc, Czech Republic}
\author{Bruno de la Torre}
\affiliation{Institute of Physics of the Czech Academy of Sciences, v.v.i., Cukrovarnick\'{a} 10, 162 00 Praha 6, Czech Republic}
\alsoaffiliation{Regional Centre of Advanced Technologies and Materials, Palack\'{y} University, \v{S}lechtitel\^{u} 27, 78371 Olomouc, Czech Republic}
\author{Mar\'{i}a Moro-Lagares}
\affiliation{Institute of Physics of the Czech Academy of Sciences, v.v.i., Cukrovarnick\'{a} 10, 162 00 Praha 6, Czech Republic}
\alsoaffiliation{Regional Centre of Advanced Technologies and Materials, Palack\'{y} University, \v{S}lechtitel\^{u} 27, 78371 Olomouc, Czech Republic}

\author{Barbora Papou\v{s}kov\'{a}}
\author{Giorgio Zoppellaro}
\affiliation{Regional Centre of Advanced Technologies and Materials, Palack\'{y} University, \v{S}lechtitel\^{u} 27, 78371 Olomouc, Czech Republic}

\author{Pingo Mutombo}
\affiliation{Institute of Physics of the Czech Academy of Sciences, v.v.i., Cukrovarnick\'{a} 10, 162 00 Praha 6, Czech Republic}

\author{Mario Ruben}
\affiliation{Karlsruhe Institute of Technology, Institute of Nanotechnology Hermann-von-Helmholtz-Platz 1, 76344 Eggenstein-Leopoldshafen, Germany}
\alsoaffiliation{Institut de Physique et Chimie des Mat\'{e}riaux de Strasbourg (IPCMS), CNRS-Universit\'{e} de Strasbourg, 67034 Strasbourg, France}

\author{Radek Zbo\v{r}il}
\affiliation{Regional Centre of Advanced Technologies and Materials, Palack\'{y} University, \v{S}lechtitel\^{u} 27, 78371 Olomouc, Czech Republic}

\author{Pavel Jelinek}
\email{jelinekp@fzu.cz}
\affiliation{Institute of Physics of the Czech Academy of Sciences, v.v.i., Cukrovarnick\'{a} 10, 162 00 Praha 6, Czech Republic}
\alsoaffiliation{Regional Centre of Advanced Technologies and Materials, Palack\'{y} University, \v{S}lechtitel\^{u} 27, 78371 Olomouc, Czech Republic}

\title{On-Surface Structural and Electronic Properties of Spontaneously Formed Tb$_2$Pc$_3$ Single Molecule Magnets}

\begin{document}

\abstract{The single molecule magnet (SMM)  bis(phthalocyaninato)terbium (III) (TbPc$_2$) has attracted steady research attention as an exemplar system for realizing molecule-based spin electronics.  In this paper, we report on the spontaneous formation of Tb$_2$Pc$_3$ species from TbPc$_2$ precursors via sublimation in ultrahigh vacuum (UHV) onto an Ag(111) surface. The molecules on the surface are inspected using combined scanning tunneling (STM) and non-contact atomic force microscopies  (nc-AFM) at 5 Kelvin.  Submolecular resolution and height dependent measurements supported by density functional theory (DFT) calculations unambiguously show the presence of both TbPc$_2$ and Tb$_2$Pc$_3$ species.  The synthesis of Tb$_2$Pc$_3$ species under UHV conditions is independently confirmed by chemical analysis.  The high-resolution AFM imaging allows us to register the orientation of the topmost Pc ligand in both Tb$_2$Pc$_3$ and TbPc$_2$  relative to the underlying Ag(111) surface.  Measurements of the electronic structure reveal the selective appearance of a Kondo signature with temperature $\sim$ 30K in the Tb$_2$Pc$_3$ species, localized to the Pc ligand lobes.  We attribute the presence of the Kondo resonance on select Tb$_2$Pc$_3$ molecules to the orientation of internal molecular ligands. High-resolution AFM imaging identifies geometric distortions between Tb$_2$Pc$_3$ molecules with and without the Kondo effect, the result of the complex interplay between structural and electronic differences.
}

\section{Introduction}

Single-molecule magnets (SMMs) are metal complexes which act as magnetic domains at the single-molecule level; \cite{Sessoli1993, Gatteschi2006, Woodruff2013} the nanosize dimensions and quantum nature of SMM systems brings to light several properties that link macroscopic phenomena with the quantum world, such as the emergence of staircase hysteretic behavior in the magnetization, quantum phase interference, and temperature independent relaxation processes. SMMs are synthesized by coordinating spin active transition-metal ions (\emph{d}, \emph{f}-block) with a variety of organic-based chelating molecules, in such a way that the unpaired electrons located on the metal ions are coupled together via exchange interactions and give rise to systems with high spin multiplicity, and large (as well as negative) zero field splitting with dominant uniaxial magnetic anisotropy. Diverse examples of multi-metal SMM clusters are described in literature, starting from the Mn12-Ac molecule, which represents one of the most studied, early SMM prototypes (U$_{eff}$ = 51 cm$^{-1}$, S = 10 and D = -0.51 cm$^{-1}$). \cite{Sessoli1993} 

Single ion molecular magnets (SIMs) and, in particular, the bis(phthalocyaninato) terbium(III) (TbPc$_2$) molecule, stands out in recent years for being extensively investigated.  Structurally, the TbPc$_2$ molecule adopts a D$_{4d}$-symmetry and a square anti-prismatic terbium coordination geometry. The complex can exist in both anionic and uncharged (neutral) form due to the redox non-innocent Pc ligand.  These very promising SIM candidates exhibit particularly high energy barriers for magnetization reversal (U$_{eff}$ = 410-641 cm$^{-1}$ depending on the Pc-ring oxidation states and environment) \cite{Woodruff2013} and blocking temperatures (T$_c$ $>$ 1.7 K); properties that are not quenched even when the molecule is subjected to dimensionally constrained environments (e.g. following surface deposition). \cite{MorenoPineda2016} These key features may promote the technological implementation of the system in real-world devices. \cite{Kuch2017, Komeda2011, LodiRizzini2011, Vincent2012, Schwobel2012, Fu2012b, Katoh2012, LodiRizzini2012, Robles2012, Thiele2014, Komeda2014b, Mullegger2014b, LodiRizzini2014, Mannini2014, Komeda2014c, Nistor2015, Robaschik2015, Zhang2015a, Perfetti2016a, Ara2016, Katoh2016, Warner2016a, Serrano2016, Wackerlin2016, Marocchi2016, Amokrane2017, Serri2017}  In particular, measurements utilizing radio frequency and magnetic field excitations in TbPc$_2$ point towards the potential use of this molecule for reading the nuclear spin state information of the Tb atom through the coupled electronic states of the metal ion and the ligands. \cite{Mullegger2014b, Urdampilleta2015, Godfrin2017, Moreno-Pineda2017}

One prevalent and widely studied phenomena on TbPc$_2$ molecules is the selective and controllable presence of a zero-bias Kondo resonance in the dI/dV spectra, localized to the Pc ligand lobes.  \cite{Katoh2009, Komeda2011, Komeda2014c, Ara2016, Warner2016a, Amokrane2017} The location and strength of the Kondo resonance has been attributed to charge transfer between the substrate and the TbPc$_2$ molecules. \cite{Vitali2008, Ara2016, Mannini2014, Marocchi2016, Thiele2014, Warner2016a, Katoh2012, LodiRizzini2011, Kuch2017, Komeda2011, Nistor2015, Wackerlin2016, LodiRizzini2014, Katoh2009, Robles2012, Zhang2015a, Komeda2014b, Komeda2014c, Amokrane2017, Katoh2016} The Kondo phenomenon is observed in all studies conducted on the monolayer TbPc$_2$ on Au(111)  \cite{Serrano2016, Komeda2014c} but not on the Ag(111) substrate. \cite{Ara2016} Nevertheless, in the case of the higher apparent height (i.e. brighter) molecules, on Au(111) the Kondo effect is quenched, but is present on Ag(111).  Additionally, the Kondo state can be quenched by physical rotation of the top Pc ligand, \cite{Komeda2011, Fu2012b} and has also been shown to be influenced by the ligand neighborhood. \cite{Amokrane2017} Spin resolved measurements demonstrate the importance of understanding exactly how interaction between the substrate and molecule, e.g., charge transfer, influences the orbital occupancy and resulting electronic properties of the molecules. \cite{Schwobel2012}

Despite these extensive research efforts, ambiguities about certain basic structural information remain for TbPc$_2$ molecules absorbed on a surface.   It has been previously observed that at elevated temperatures, TbPc$_2$ molecules may decompose and form Tb$_2$Pc$_3$ and Tb$_3$Pc$_4$, with ratios that seem to depend on the experimental conditions such as annealing temperatures employed and use of an UHV environment. \cite{Stepanow2010} However the systematic study of such species on surfaces has been missing so far, to our knowledge.  Moreover, the possibility to form  triple-decker Tb$_2$Pc$_3$ and in general Ln$_2$Pc$_3$ SMMs seems to be an unlikely scenario for synthetic chemistry in solution.  Indeed, little is known about triple-decker Tb$_2$Pc$_3$ and in general Ln$_2$Pc$_3$ SMMs, because the unsubstituted phthalocyanine rings render such systems difficult to handle in any solvent. Thus, only few multi-decker Ln complexes are known today in literature, and mostly comprise of species making use of substituted Pc-rings (e.g. with OEt, OBu) to increase solubility. \cite{Woodruff2013}

Here we report measurements of TbPc$_2$ molecules thermally evaporated in ultrahigh vacuum (UHV) on a single crystal Ag(111) surface and subsequently studied via scanning tunneling (STM) and non-contact atomic force microscopies  (nc-AFM).  \cite{Jelinek2017}  Submolecular spatial AFM resolution achieved with a CO-terminated probe allows us to identify the presence of well ordered, self-assembled islands composed of both TbPc$_2$ and Tb$_2$Pc$_3$ molecules.  The unexpected presence of Tb$_2$Pc$_3$ molecules is further supported by manipulation experiments and confirmed by chemical analysis of the molecules collected after the UHV experiments.  From these observations we can infer an alternating rotational configuration of the triple Pc ligand stack, which can rationalise the appearance of the Kondo resonance observed on Tb$_2$Pc$_3$  molecules. The experimental evidence is supported by total energy density functional theory (DFT) calculations.

\section{Results and Discussion}

\begin{figure}
\centering
\includegraphics[keepaspectratio=true, width=.5\linewidth]{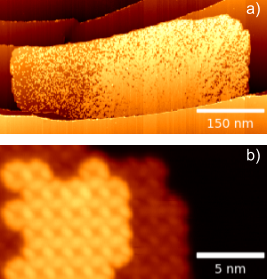}
\caption{a) Overview STM topography (V$_{bias}$ = 900 mV, I = 10 pA) showing a close-packed complete molecular island. b) Close-up STM (V$_{bias}$ = 900 mV, I = 15 pA) from island edge showing two heights of molecule and bare Ag(111).}
\label{overview-fig}
\end{figure}

After deposition of  TbPc$_2$ molecules on the Ag(111) surface held at room temperature  (for details see Methods), we observed in STM large areas ($>150 \times 400$ nm) of densely packed molecules surrounded by clean Ag(111), as shown in Fig. \ref{overview-fig}a). The molecular layer grows between surface step edges and we clearly observe two different apparent height contrasts within the molecular layer using STM imaging. Detailed inspection of the edge of an island as shown in Fig. \ref{overview-fig}b), reveals two distinct molecular contrasts, dark and bright, which correspond to different apparent heights of molecules. Differences in height have been previously observed in STM measurements for TbPc$_2$ absorbed on Au(111) and Ag(111) surfaces \cite{Komeda2011, Ara2016} and additional species with higher apparent height were attributed to a second layer of TbPc$_2$ molecules. \cite{Ara2016, Serrano2016}

\begin{figure}
\centering
\includegraphics[keepaspectratio=true, width=1\linewidth]{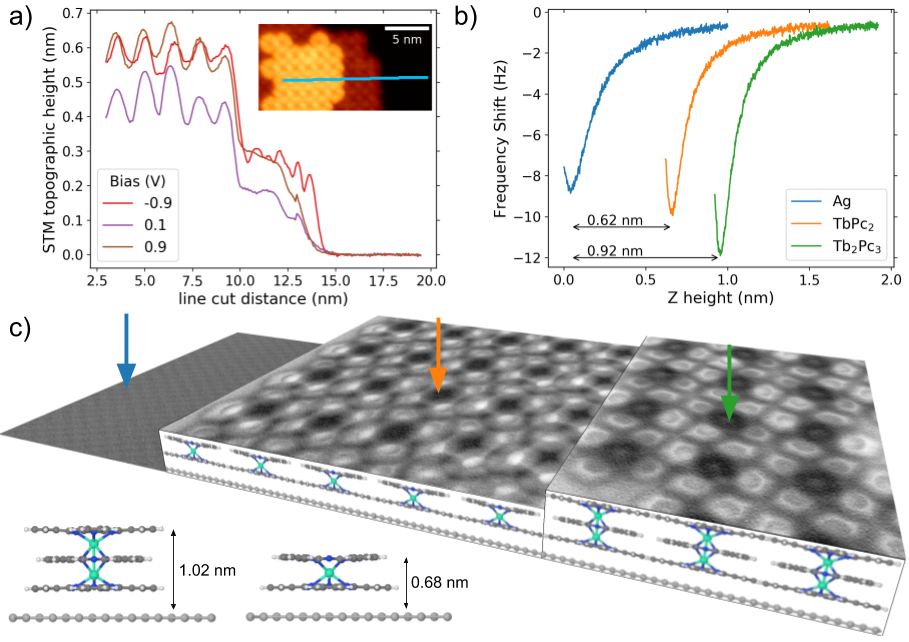}
\caption{Comparison of STM and AFM height measurements.  a) Bias-dependent line profiles of inset STM topography (V$_{bias}$ = 900 mV, I = 15 pA) showing Tb$_2$Pc$_3$, TbPc$_2$ and bare Ag(111) surface.  b) $\Delta f$(z) spectroscopies taken on the three different regions indicated by the colored arrows on c).  We use the minima from these curves to determine the heights of TbPc$_2$ and Tb$_2$Pc$_3$ as 0.62 nm and 0.92 nm, respectively.  c) Schematic diagram showing multi-height high-resolution imaging of the Ag(111), TbPc$_2$ and Tb$_2$Pc$_3$ molecules.  The DFT optimized models have heights of 0.68 and 1.02 nm, shown in the lower left corner.}
\label{STM-AFM-height}
\end{figure}

However, apparent topography measured in STM on the same region may change substantially at different tip-sample biases.  Indeed, bias-dependent line profiles, presented in Fig. \ref{STM-AFM-height}a), show significant deviation in the apparent STM height between the two different species, making reliable measurement of the height impossible using this method.  For certain bias voltages, the higher apparent height molecules are roughly double that of the lower apparent height molecules relative to the Ag(111) surface, which is likely why they have previously been identified as a second layer of TbPc$_2$ molecules. \cite{Ara2016, Serrano2016}

\textbf{STM and AFM: height determination}

To overcome this deficiency in STM height determination, we examined the apparent heights in the nc-AFM mode using a CO-terminated tip. The presence of an inert CO molecule on the metallic tip apex significantly reduces the chemical reactivity of the probe, which allows us to reach the frequency shift minima without variation of tip structure.  This measurement has been previously demonstrated to be an accurate way to determine the heights of molecules absorbed on surfaces. \cite{Schuler2013}  The frequency shift was recorded as a function of tip-sample distance on the two different molecular regions and Ag(111) surface, as shown in Fig. \ref{STM-AFM-height}b).  By identifying the frequency shift minimum on the three regions (dark, bright molecules and bare Ag(111) surface), we obtained the height of the two different molecular regions to be 0.62 nm and 0.92 nm with respect to the Ag(111) surface. In contrast with the apparent height from STM, these measurements show that the higher molecules are not twice as high as their lower counterparts. On the other hand, the two AFM determined heights match well with the distances between the upper Pc ligand of TbPc$_2$ and Tb$_2$Pc$_3$, with respect to the Ag(111) surface as obtained from the total energy DFT simulations, shown in Fig. \ref{STM-AFM-height}c). This finding strongly suggests that the higher and lower molecules in STM represent Tb$_2$Pc$_3$ and TbPc$_2$  molecules, respectively.

\begin{figure}
\centering
\includegraphics[keepaspectratio=true, width=.9\linewidth]{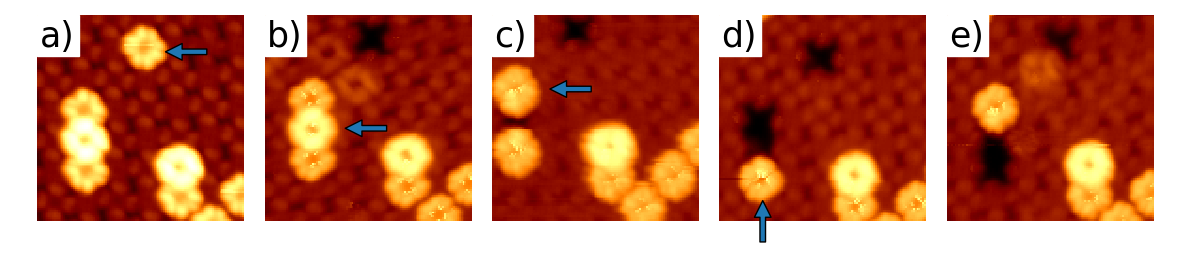}
\caption{Manipulation experiments performed on Tb$_2$Pc$_3$ molecules, indicated by the blue arrows.  a) Initial state of the region.  b) When extracted (absorbed to the tip), hole down to bare Ag surface becomes visible.  c) and d) show the removal of additional Tb$_2$Pc$_3$ species to create a channel; d) and e) demonstrate how the remaining Tb$_2$Pc$_3$ moves laterally in the channel as a discrete unit.}
\label{manipulation-fig}
\end{figure}

\textbf{Molecular Manipulation of Tb$_2$Pc$_3$}

We performed manipulation experiments of the bright molecules, the results of which further reinforce the identification of Tb$_2$Pc$_3$ on Ag(111). Figure \ref{manipulation-fig} shows a sequence of molecular manipulations performed by moving the tip in constant height laterally into the side of different bright molecules, with the direction of tip motion indicated by the arrows.  We expected lateral manipulations to move what we initially supposed to be second layer TbPc$_2$ molecules.  Instead, for the lateral manipulations in Fig. \ref{manipulation-fig}a), \ref{manipulation-fig}b), and \ref{manipulation-fig}c), the bright molecule was completely removed from the area (e.g., absorbed to the tip), leaving a hole down to the substrate.  The removal of the molecules in Fig. \ref{manipulation-fig}b) and \ref{manipulation-fig}c) created a channel that was required for the successful lateral movement of the higher molecule, shown in Fig. \ref{manipulation-fig}d). We interpret these observations as evidence that the species with higher apparent height in STM are chemically discrete units, consistent with our identification of them as Tb$_2$Pc$_3$. 

\textbf{\emph{Ex-situ} Chemical Analysis}

To gain more evidence about the formation of Tb$_2$Pc$_3$ molecules, we performed \emph{ex-situ} chemical analysis of the molecules before and after several thermal cycles of evaporation in UHV onto the Ag(111) surface.  High-resolution mass spectrometry (time-of-flight tandem mass spectrometer, Q-TOF) for residual material remaining in the crucible after the UHV experiments, confirms the presence of both the TbPc$_2$ and Tb$_2$Pc$_3$ species  (ESI material, [TbPc$_2$]$^+$ with 1183.2 m/z$^+$ and [Tb$_2$Pc$_3$]$^+$ with 1855.3 m/z$^+$) as well as the presence of increasing signatures attributable to half-decker Tb molecules [TbPc]$^+$ (see the Supplementary Material). From these observations we conclude that thermal decomposition in the UHV chamber appears to promote a temperature-induced structural rearrangement of the TbPc$_2$ molecules, similar to what has been shown earlier, \cite{Stepanow2010} following a reaction pathway tentatively highlighted by the following scheme: $4\mathrm{TbPc}_2 (\Delta T) \rightarrow \mathrm{Tb}_2\mathrm{Pc}_3 + \mathrm{TbPc}_2 + \mathrm{TbPc} + 2 \mathrm{Pc}$.

\textbf{Structural Model}

We performed detailed measurements and analysis of the arrangement of TbPc$_2$ and Tb$_2$Pc$_3$ assemblies on the Ag(111) surface.  The nc-AFM technique provides unprecedented spatial resolution of molecular assemblies superior to standard STM imaging. \cite{Jelinek2017} In particular, the technique allows us to precisely measure the internal orientation of molecules forming the on-surface assembly.

\begin{figure}
\centering
\includegraphics[keepaspectratio=true, width=1\linewidth]{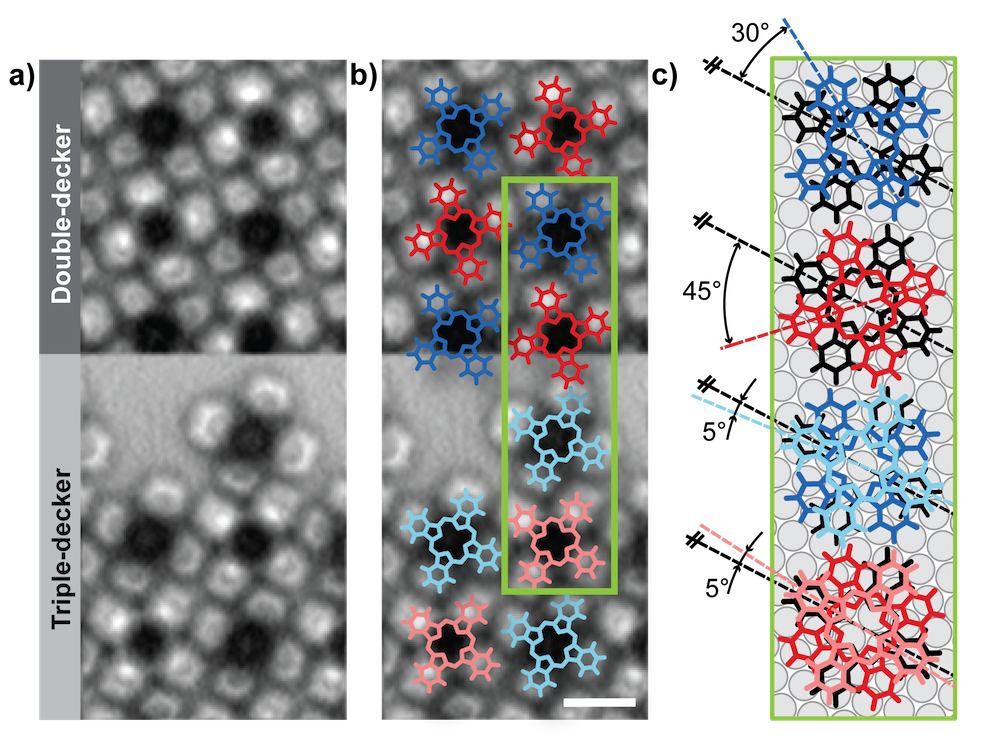}
\caption{nc-AFM registration of top Pc ligand and absorption model.  a) Multi-height nc-AFM resolving the top ligand of the TbPc$_2$ and Tb$_2$Pc$_3$ areas.  b) Orientation of the double-decker TbPc$_2$ and triple-decker Tb$_2$Pc$_3$ is shown with a superimposed model of the top Pc ligands.  White scale bar is 1nm.  c) Model showing the absorption sites of Pc on Ag(111) (black), and rotational orientations of the subsequent ligands.}
\label{model-fig}
\end{figure}

Figure \ref{model-fig} shows the multiple-height registration of the TbPc$_2$ relative to the Tb$_2$Pc$_3$ deduced from analysis of a single high-resolution AFM image of two domains of TbPc$_2$ and Tb$_2$Pc$_3$ molecules.  We have directly measured the position and rotational orientation of the topmost Pc in both the double- and triple-decker species with respect to the Ag(111) lattice (more data shown in the Supplementary Material, Figs. S1 and S2).  We inferred the adsorption geometry of the bottom Pc ligand (black, Fig. \ref{model-fig}c) from the atomic registration, and an equidistant ligand spacing due to steric repulsion, consistent with prior observations. \cite{Bai2008}  For the TbPc$_2$ molecules, we can directly add the topmost ligand from the experimental observations, with an alternating 30$^\circ$ / 45$^\circ$ pattern relative to the bottom Pc (dark blue/red in Fig. \ref{model-fig}b).  For the Tb$_2$Pc$_3$, we infer that the middle Pc conforms to the alternating 30$^\circ$ / 45$^\circ$ pattern of the topmost TbPc$_2$, and then add the topmost Pc, with a $\pm 5^\circ$ rotation relative to the bottom Pc (light blue/red).

From these observations, we can identify a two-molecule unit cell commensurate with the Ag(111) lattice (shown in supplementary information Fig. S2).  Unfortunately, such a big unit cell makes total energy DFT simulations computationally intractable.  We performed calculations in a smaller unit cell including only one TbPc$_2$ or Tb$_2$Pc$_3$ species, respectively.  This means that the mutual orientation cannot be directly compared to the experimental evidence. According to the total energy DFT simulations the TbPc$_2$ or Tb$_2$Pc$_3$ molecules are stabilised on the Ag(111) surface by an attractive van der Waals interaction and charge transfer between the molecular species and metallic substrate. In both cases, we observed substantial planarization of the Pc ligands with respect to the optimal gas phase configuration (see inset in Fig. \ref{STM-AFM-height}c)), as a consequence of the attractive van der Waals interaction with the metallic substrate. The planar structure of the Pc ligands facilitates sub-molecular resolution measurements. Indeed, the calculated AFM images are consistent with the experimental data (see Supplementary Material Fig. S5).

\textbf{Molecular Charge Transfer}

\begin{figure}
\centering
\includegraphics[keepaspectratio=true, width=.8\linewidth]{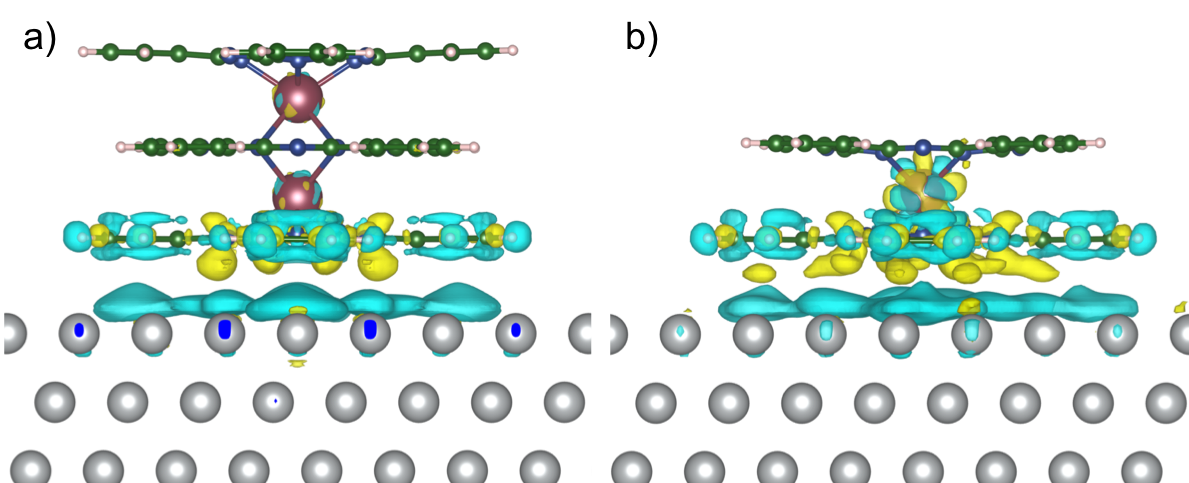}
\caption{Calculated charge transfer between Tb$_2$Pc$_3$ and TbPc$_2$ molecules and Ag(111) surface obtained from DFT calculations.  The charge transfer is represented by differential charge $\rho_{ALL}$ - $\rho_{mol}$ - $\rho_{surf}$, where $\rho_{ALL}$ is total density of the whole system and $\rho_{mol}$ and $\rho_{surf}$ are calculated densities of molecule and Ag(111) surface respectively. The yellow and blue colours represent accumulation and loss of density, respectively. The presence of blue density on the upper surface layer indicates substantial charge transfer from the metallic surface towards the molecule. The presence of both blue and yellow differential densities on the molecule reveals charge redistribution within the molecule upon charge transfer.}
\label{DFT-fig}
\end{figure}

We extended the nc-AFM measurements to look for differences between the two observed rotational configurations, specifically a detailed examination of the charge distribution across molecule/surface interface and within each molecule.  To analyze the charge transfer, we carried out  Kelvin probe force microscopy (KPFM) measurements on the two species of molecules as well as the bare Ag(111) (see Supplementary Material Fig. S4).  Differences in the capacitive term and local contact potential difference were observed between the three regions.  We found the magnitudes of these parameters to be strongly dependent on the tip geometry and tip-sample distance, but the overall trend between Tb$_2$Pc$_3$, TbPc$_2$ molecules and the Ag surface remained very similar.  We observed differences in the contact potential difference V$_{CPD}$ between the molecule regions and the bare Ag(111), indicating that there is charge transfer occurring from the substrate towards the molecules, which is also accompanied by charge transfer between domains of  Tb$_2$Pc$_3$, TbPc$_2$ molecules. The charge transfer from the surface towards molecules is further supported by total energy DFT simulations of single TbPc$_2$ and Tb$_2$Pc$_3$ molecules deposited on Ag(111) surface, as shown on Fig. \ref{DFT-fig}. Namely, the DFT calculations indicates an extensive depletion of the electron density in surface area (see blue color iso densities shown on Fig. \ref{DFT-fig}) under the molecule. On the other hand, we observe substantial charge redistribution in lower Pc and on Tb atom. This we attribute to molecular orbital reordering upon deposition on the Ag(111) surface. Our simulations predict that upon surface deposition the eight $f$-electrons of Tb still remain lined up with a total spin S = 3 and a total orbital momentum L = 3. In the next section, we will show that the charge transfer plays an important role in the appearance of the Kondo effect.

Unfortunately, we do not observe differences on submolecular level between adjacent Tb$_2$Pc$_3$ molecules in the KPFM data. Nevertheless this level of spatial resolution for KPFM is likely limited by mesoscopic tip effects. In principle, we can improve it with an intentional CO-tip functionalization. But to obtain high spatial resolution, we have to operate the probe in close distances and large biases, which gives rise to spurious artefacts in the KPFM measurements. \cite{Albrecht2015}

\begin{figure}
\centering
\includegraphics[keepaspectratio=true, width=.5\linewidth]{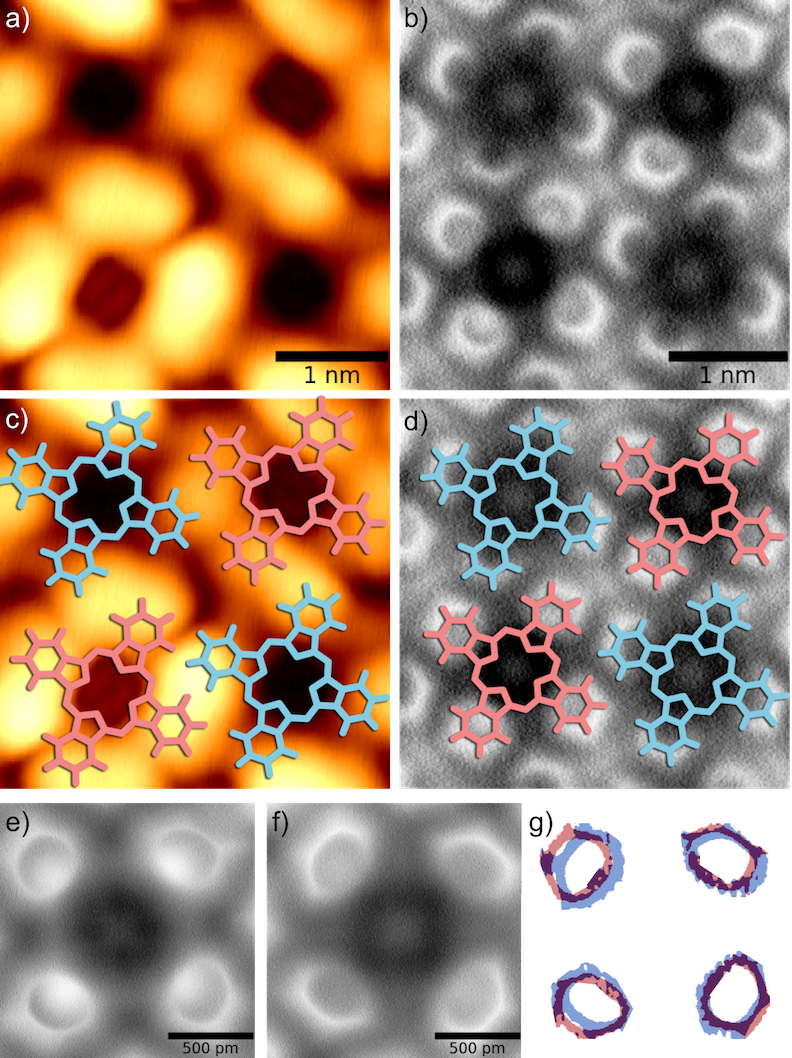}
\caption{Comparison of equivalent Tb$_2$Pc$_3$ regions with STM and nc-AFM imaging.  a) STM image measured on four molecules, V$_{bias}$ = 50 mV, I = 10 pA. b) nc-AFM image taken in constant height mode, recording the frequency shift in space over the same region as a). c) and d) have the top Pc ligand model overlaid to guide the eye.  e) and f) are detailed images of the upper right- and left-hand molecules in b), respectively.  They are subsequently registered together and plotted in red (e)/blue (f) in g).}
\label{HR-fig}
\end{figure}

Because of the limited spatial KPFM resolution we pursued a different approach to address differences in the internal charge distribution within molecules.  Molecular backbones are typically represented by sharp edges caused by lateral bending of the CO-tip.  Modification of the internal charge distribution within the molecule modifies the Coulombic interaction experienced by CO-tip. This in turn modifies the position of sharp edges providing information about the variation of the charge distribution. \cite{Hapala2016}  We applied this methodology to this system to see if we could correlate the structural and electronic differences we observed in the sub-molecular AFM contrast.  Figures \ref{HR-fig}a) and \ref{HR-fig}b) are images of the same group of four Tb$_2$Pc$_3$ molecules, taken in standard STM with a metallic tip (Fig. \ref{HR-fig}a), and nc-AFM at constant height with a CO- functionalized probe (Fig. \ref{HR-fig}b).  The top Pc ligand model is superimposed on Figs. \ref{HR-fig}c) and \ref{HR-fig}d) to guide the eye.  The upper left- and right-hand molecules in Fig. \ref{HR-fig}b) were imaged in greater detail, shown in Figs. \ref{HR-fig}f) and \ref{HR-fig}e) respectively.  By filtering and registering these two images to one another, slightly different distortions in the benzene rings can be identified, as shown in Fig. \ref{HR-fig}g). Namely, we observe a systematic elongation of benzene rings corresponding to red molecules (two lower Pc ligands are rotated by 45$^\circ$ with respect to each other, see  Fig. \ref{model-fig}c).  Similar geometric distortions have been attributed to charge transfer effects, owing to the combination of Pauli repulsion, van der Waals and Coulombic interaction of the CO-tip with the sample. \cite{Hapala2014, Hapala2016, DeLaTorre2017, Peng2018} Therefore, we attribute these differences in lateral distortion of benzene rings to different charge distributions in adjacent Tb$_2$Pc$_3$ molecules.  Unfortunately, due to the lack of submolecular resolution of the internal part of the molecules, a direct comparison of the electrostatic potential between two Tb$_2$Pc$_3$ molecules with different internal orientations of Pc ligands cannot be accomplished. 

It is noteworthy that very similar alternating distortions were also observed in adjacent TbPc$_2$ molecules with different internal rotations of Pc ligands. This indicates that the different rotations of lower and upper Pc ligands modifies the hybridization with the central Tb atoms and consequently modifies the charge distribution within the molecule.

\textbf{Electronic Structure of TbPc$_2$ and Tb$_2$Pc$_3$}

\begin{figure}
\centering
\includegraphics[keepaspectratio=true, width=.5\linewidth]{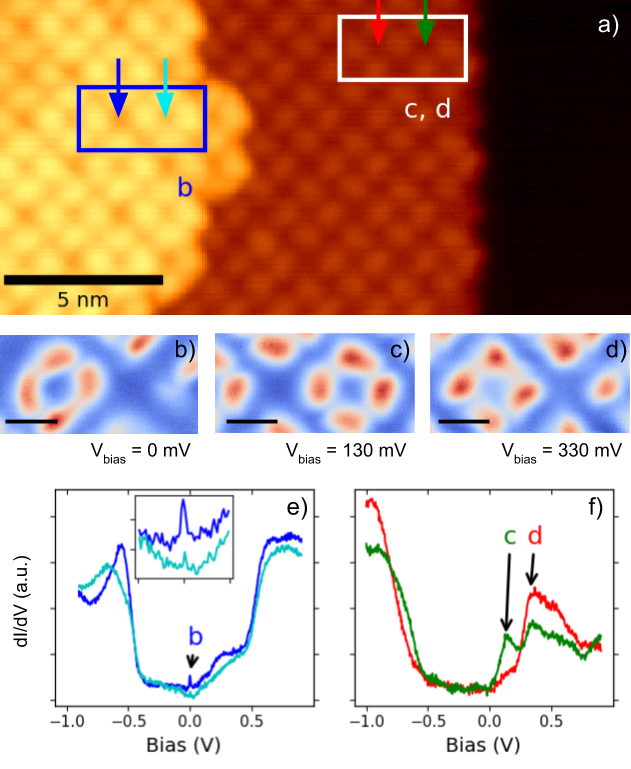}
\caption{Characterization of the electronic structure of Tb$_2$Pc$_3$ and TbPc$_2$ areas.  a) STM topography (V$_{bias}$ = 300 mV, I = 10 pA) showing the locations for dI/dV spectroscopies averaged over the ligand (excluding the center) measured on adjacent molecules in the Tb$_2$Pc$_3$ (e) and TbPc$_2$ (f) layers.  Constant height dI/dV mapping of the Kondo feature on the Tb$_2$Pc$_3$ layer (V$_{bias}$ = 0) is shown in b).  Analogous maps of the TbPc$_2$ region at V$_{bias}$ = 130 mV (c) and 330 mV (d).  Scale bars for b), c), and d) are 1 nm.  Inset in e) shows dI/dV for the bias range -0.1 to 0.1 V.}
\label{STS-map}
\end{figure}

Upon identifying the two distinct geometries of TbPc$_2$ or Tb$_2$Pc$_3$ in the surface assembly, we characterized their respective electronic properties.  We carried out scanning tunneling spectroscopic (STS) measurements of the electronic structure, which show qualitative differences between distinctly orientated molecules of each species.  Averaged STS spectra acquired on the ligand ring of two adjacent Tb$_2$Pc$_3$ and TbPc$_2$ molecules are plotted on Figs. \ref{STS-map}e) and \ref{STS-map}f), respectively.  We clearly observe distinct positions of the highest occupied/ lowest unoccupied molecular orbital (HOMO/LUMO) resonances for Tb$_2$Pc$_3$ and TbPc$_2$ molecules, which are spatially localized on the ligand. In the case of TbPc$_2$ molecules HOMO resonances are located $\sim$ 0.2 eV lower in energy than those of Tb$_2$Pc$_3$. We can also identify that the width of the HOMO peaks are directly linked to the internal rotation of Pc ligands, see  Fig. \ref{STS-map}e). The lowest unoccupied states are spatially localized on Pc ligands similarly to the HOMO orbitals, as seen from spatial STS maps of two adjacent TbPc$_2$ with alternating relative Pc orientations at V$_{bias}$ = 130mV and 330mV shown in Fig. \ref{STS-map}c) and \ref{STS-map}d), respectively. 

The most significant observation from these data is the presence of the zero bias anomaly in the STS spectra only on one type of Tb$_2$Pc$_3$ molecule, while neither the second type of Tb$_2$Pc$_3$ nor TbPc$_2$ molecules display any evidence of the resonance. The alternating presence of the zero bias anomaly is also demonstrated explicitly in the real-space STS map at the Fermi level (V$_{bias}$ = 0 V), shown in Fig. \ref{STS-map}b).  The presence of the Kondo effect in TbPc$_2$ molecules deposited on different metal surface has already been reported by many groups. \cite{Warner2016a, Komeda2014c, Serrano2016, Katoh2009, Katoh2016, Komeda2011, Amokrane2017, Komeda2014b, Schwobel2012, Ara2016, Katoh2012}  Consequently we attributed this zero bias anomaly to the Kondo resonance.  Our spectroscopic measurements establish that the Kondo resonance appears only for the Tb$_2$Pc$_3$ molecules having the HOMO orbital located closest to the Fermi level.  From the previous structural analysis, we can assign the presence of the Kondo signature to the molecules overlaid with the light red model shown in Fig. \ref{model-fig}b), corresponding to the middle ligand having a 45$^\circ$ rotation relative to the bottom.

\begin{figure}
\centering
\includegraphics[keepaspectratio=true, width=.5\linewidth]{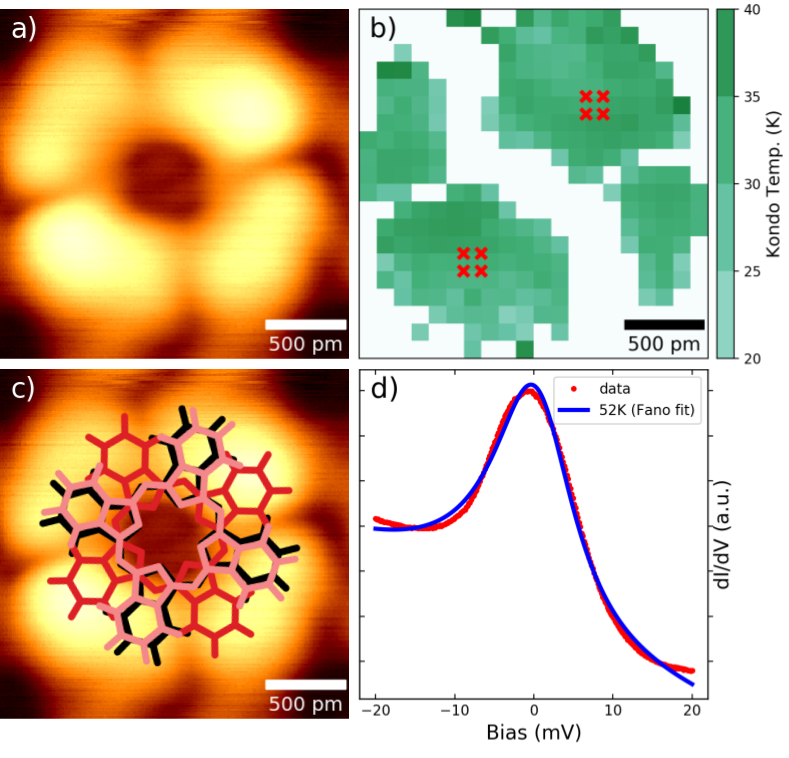}
\caption{Spatial mapping of the Kondo peak.  a) STM topography (V$_{bias}$=50 mV, I = 10 pA) of the molecule subsequently mapped with grid spectroscopy.  b) Map of the fitted Kondo temperature on a single Tb$_2$Pc$_3$ species, showing a consistent temperature of approximately 30K (Frota procedure) located on the Pc ligands.  c) Topography from a) with Pc ligand superimposed to guide the eye.  d) Averaged point spectra showing the Kondo peak (x's on b; red points), and subsequently fit with the Fano procedure (blue).}
\label{Kondo-map}
\end{figure}

To estimate the Kondo temperature $T_K$ of the Tb$_2$Pc$_3$ molecule showing the Kondo effect, we fit selected spectra using the Frota expression for the Kondo resonance: \cite{Frota1992, Pruser2011}

\begin{equation}
\rho_F(\epsilon) = Im \left[-i \sqrt{\frac{i \Gamma_F}{i \Gamma_F + \epsilon}} \right],
\end{equation}
with an additional linear background offset, where $\epsilon$ is the energy and $\Gamma_F$ is proportional to the half width at half maximum.  The Kondo temperature $T_K$ can be calculated from $\Gamma_F$ as follows: \cite{Zitko2009, Pruser2011} $\Gamma_F = 1.455 k_B T_{K}$, where $k_B$ is the Boltzmann constant.  These fits indicate a temperature of approximately 30 K ($\pm 5$ K), consistent with prior measurements. \cite{Ara2016}  The Kondo temperature $T_K$ has been averaged for several spectra acquired on different lobes of the upper Pc ligand, in positions indicated by red points in Fig. \ref{Kondo-map}b). There can be only one Kondo temperature for a given molecular state.  We attribute the minor spatial variation of the Kondo resonance to vagaries in the tunneling process of electrons between tip and molecular states in the given energy window.  Consequently, we intentionally selected spectra in the central part of ligand, which are less affected by the convolution process. The estimated Kondo temperature is far above the experimental temperature $T_{exp}$ = 5K, which indicates that the system is in the strongly coupled Kondo regime.

Next, we analyze the Kondo effect in detail to gain further insight regarding its selective appearance on only the Tb$_2$Pc$_3$ molecules with the middle ligand rotated by 45$^\circ$ with respect to the bottom ligand.  The presence of the Kondo resonance is intimately connected to the interaction of singly occupied magnetic impurity states with a bath of non magnetic electrons, typically located in a metallic substrate. This interaction leads to an effective screening of the spin of the localized state, which is manifested by presence of a sharp resonance in the spectral function at the Fermi level. \cite{Hewson1993}  The width of the Kondo resonance $\Gamma$ is proportional to the Kondo temperature $T_K$, which defines the relevant energy scale $k_{B} \Gamma$ of the Kondo effect.  The Kondo temperature is determined by several factors, such as the position of the singly occupied state $\epsilon_i$ with respect to the Fermi level, its hybridisation with the substrate $\Delta$ and on-site Coulomb interaction $U$: \cite{Hewson1993}

\begin{equation}
\Gamma = k_B T_K  \approx \sqrt{2 \Delta \frac{U}{\pi}} \mathrm{exp} \left[ - \frac{\pi}{2 \Delta} \left(\left|\frac{1}{\epsilon_i}\right| + \left|\frac{1}{\epsilon_i + U}\right|\right)^{-1} \right].
\label{Hewson-eqn}
\end{equation}

Another parameter that plays an important role in the quantitative analysis of the Kondo problem is an average occupancy number $n_i$ of the impurity state $\epsilon_i$. \cite{Hewson1993} In our case, it is represented by the highest occupied molecular level. It has been demonstrated that the Kondo regime is typically established within an occupancy range of 0.8 $< n_i <$ 1.2 having approximately one unpaired electron in the highest occupied molecular level. \cite{Zlatic1985} In the gas phase this is an integer number, but this is not necessarily the case upon deposition on a metallic surface due to additional interaction with the substrate. Indeed, we already established that both KPFM measurements and DFT calculations show charge transfer towards the molecule accompanied by additional charge transfer between domains of TbPc$_2$ and Tb$_2$Pc$_3$.  The high resolution AFM images reveal internal charge redistribution of the electrostatic field within molecules and the position of the HOMO with respect to the Fermi level induced by internal rotation of the Pc ligands.

We can estimate the occupancy $n_i$ of the HOMO orbital of Tb$_2$Pc$_3$ molecules showing the Kondo resonance from analysis of the STS spectra using the well known Fano formula: \cite{Fano1961}

\begin{equation}
\rho(E) = \rho_0 + \frac{(q + \epsilon)^2}{1+\epsilon^2},
\end{equation}

with an additional linear offset term, where $\epsilon$ is the normalized energy:

\begin{equation}
\epsilon = \frac{E - E_K}{\Gamma_{exp}}.
\end{equation}

$E_K$ is the peak position, and $\Gamma_{exp}$ is the half-width at half-maximum of the STS spectrum.  $q$ is the Fano parameter which interpolates between a Lorentzian peak ($q \rightarrow \infty$) and dip ($q = 0$), depending on the strength of the interference effect between competing tunneling channels through the single impurity and its environment.  In our case, we obtained the best fit with parameters $E_K$ = 0.62 meV, $\Gamma_{exp}$ = 6.3 meV (blue line, Fig. \ref{Kondo-map}d).  We correct the $\Gamma$ width parameter for the thermal and electronic broadening by considering their root-mean-square contributions to the effective and intrinsic peak width: \cite{Ara2016}

\begin{equation}
\Gamma_{exp} = \sqrt{(5.4 \frac{k_b T}{e})^2 + (1.7 V_{rms})^2 + \Gamma^2},
\end{equation}

where we take T = 5K as the temperature of the microscope, and $V_{rms}$ = 5 mV as the lock-in signal amplitude for these measurements. The corrected value $\Gamma$ = 4.4 meV gives the Kondo temperature  $T_K$ = 51K. This value is a little bit higher then the value obtained by the Frota fit.  Finally, the occupation number $n_i$ of the HOMO orbital can be deduced from the corrected $\Gamma$ using the following expression: \cite{Hewson1993}

\begin{equation}
E_K = \Gamma \, \mathrm{tan}\left(\frac{\pi}{2}(1 - n_i)\right).
\end{equation}

We obtained the average occupation $n_f$ = 0.91, which fits into the predicted range of occupancies where the Kondo effects may appear.  We can check the consistency of this fit, by using the corrected value of $\Gamma$ = 4.4 meV and solving Eqn. \ref{Hewson-eqn} subject to the additional constraint that $n_i = -\frac{\epsilon_i}{U} + \frac{1}{2}$. \cite{Ujsaghy2000} With this constraint, we obtained numerically optimised values for the Coulomb repulsion $U$ = 1.1 eV;  $\epsilon_i = -470$ meV, the distance of the HOMO below the Fermi level; and $\Delta = 105$ meV, the width of the HOMO orbital. These values reasonably match those from the experiment, which can be estimated from the STS spectra of the Tb$_2$Pc$_3$ molecule with Kondo resonance shown on Fig. \ref{STS-map}e): $U \sim$ 1.1 eV, $\epsilon_i \sim -550$ meV, and $\Delta \sim 150$ meV.  In contrast, for the non-Kondo HOMO orbital (light blue, \ref{STS-map}e) we estimate $\epsilon_i \sim -650$ meV, and for TbPc$_2$ $\epsilon_i \sim -950$ meV.

Based on the experimental evidence showing correlation between the position of HOMO orbitals with respect to the Fermi level and appearance of  the Kondo resonance, we attribute the appearance of Kondo resonances to two factors: (i) the charge transfer from the surface to molecules and between different domains of TbPc$_2$ and Tb$_2$Pc$_3$ molecules, and (ii) the internal rotation of Pc ligands, further modifying the internal charge redistribution within molecules and importantly modifying the position of the highest occupied molecular resonance. Based on these observations, we propose the following scenario to rationalize the presence/absence of Kondo resonances in Tb$_2$Pc$_3$ and TbPc$_2$ molecules.

In the case of TbPc$_2$ molecules, it has been shown that the Kondo resonance observed on TbPc$_2$ molecules on Au(111) surface is related to the presence of an unpaired electron in the $\pi$-orbital of the upper Pc ligand, as discussed by Komeda et. al. \cite{Komeda2011} However, in the case of the Ag(111) surface, the presence of a semi-occupied molecular orbital (SOMO) is suppressed by additional charge transfer upon deposition of the molecule on a metallic surface. This changes the occupancy of the HOMO and shifts its energy far from the Fermi level,  and consequently the Kondo effect is quenched.

The situation is slightly different in the case of Tb$_2$Pc$_3$ molecules, where the three Pc$^{-2}$ ligands accumulate a total charge of -6, which is fully compensated by charge +6 provided from the two Tb$^{+3}$. However, this neutrality is broken upon additional charge transfer towards the molecule after adsorption onto the surface, giving rise to the presence of an unpaired electron in the upper Pc ligand. In the case of the Tb$_2$Pc$_3$, the presence of the Kondo resonance is further modulated by the internal rotation of the Pc ligand following the scenario proposed by Komeda et. al.  \cite{Komeda2011}. Namely, the rotation of the Pc ligands changes the hybridization of the HOMO orbital, which shifts downwards its energy and increases its width. In principle, this effect increases the orbital occupancy of the HOMO level and reduces the effective exchange interaction between the localized electron in the HOMO and the bath of electrons. Consequently, the Kondo resonance disappears.

\section{Conclusion}
In summary, we have employed nc-AFM based techniques to identify our multiple height system as a mixture of TbPc$_2$ and Tb$_2$Pc$_3$ molecules, subsequently confirmed with separate chemical analysis.  We established a structural model of the mixed Tb$_2$Pc$_3$ and TbPc$_2$ monolayer relative to the underlying Ag(111) surface using high-resolution AFM imaging. Spectroscopic measurements show the selective presence of a Kondo peak in the top ligand of alternating Tb$_2$Pc$_3$ species.  High-resolution imaging of molecules exhibiting and lacking the Kondo peak show geometric distortions in the Pc ligand structure that we attribute to structural and charge environment differences.  We use these structural measurements to model and simulate the charge transfer between the substrate in the various configurations:  these quantitative differences are correlated with the qualitative appearance of the Kondo electronic signature.  These data provide a comprehensive portrait for understanding the roles of structure and charge transfer in controlling the Kondo signature, and points to their utility in evaluating future candidate SMMs for use in surface based devices.

\section{Methods}

\textbf{Experimental:} Synthesis is accomplished by direct cyclization of the ring precursor (1,2-dicyanobenzene) at high temperatures in the presence of a metal salt (e.g. Tb(acac)$_3$), 1,8-Diazabicyclo(5.4.0)undec-7-ene (DBU) and high boiling solvents (e.g. penthanol, hexanol). \cite{Branzoli2009, Stepanow2010}  The measurements described here were performed using commercial, ultrahigh vacuum (UHV) low-temperature microscopes with combined STM/nc-AFM capabilities (Specs-JT Kolibri: f$_0 \sim$ 1 MHz, Q $\sim$ 120k, K $\sim$ 540 kN/m, 50 pm amplitude and Createc-qPlus: f$_0 \sim$ 30 kHz, Q $\sim$ 17k, K $\sim$ 1.8 kN/m, 50 pm amplitude).  The Ag(111) single crystal (Mateck) was prepared by repeated Ar sputtering / annealing cycles.  The TbPc$_2$ molecules were degassed in vacuum by repeated cycling to the evaporation temperature, roughly estimated as 850K ($\pm 100$K).  Coverages were obtained with a 15 min deposition in ambient pressure of $5 \times 10^{-8}$ mbar (base pressure $1\times 10^{-10}$ mbar) with the Ag(111) surface at 300K (line-of-sight distance 30 cm), before subsequent insertion into the low-temperature microscope (5K).  CO was dosed onto the surface (pressure $8 \times 10^{-8}$ mbar, 15 seconds) after the sample temperature was $<10$K, for subsequent tip functionalization.  STS point spectra and maps were acquired using the lock-in measurement technique, with an AC signal ($\sim$1 kHz) amplitude 5 mV (point/grid spectroscopies) or 15 mV (STS maps) added to the tip-sample junction.  The presence of additional Tb$_2$Pc$_3$ molecules in the precursor TbPc$_2$ molecule batch that was annealed in UHV for the experiments was subsequently confirmed with matrix-assisted laser desorption/ionization (MALDI) spectroscopy (spectra shown in the supplementary information).

\textbf{Computational:} 
The optimized structures of TbPc$_2$ and Tb$_2$Pc$_3$ molecules molecules on Ag(111)  were calculated by the FHI-AIMS program package \cite{Blum2009} based on ab initio density functional theory (DFT).  We used exchange correlation functional PBE+U \cite{Perdew1996, Dudarev1998} with U=5eV for $f$-electrons of Tb and van der Waals interaction was approximated by the Tkatchenko-Scheffler dispersion correction method \cite{Tkatchenko2009}.  The AFM images were calculated with Probe Particle code \cite{Hapala2014b, Hapala2014}.  We used the following parameters of the flexible probe-particle tip model: the effective lateral stiffness $k$ = 0.24 N/m and effective atomic radius R$_c$ = 1.661 \AA. We added a quadrupole-like charge distribution at the tip apex to simulate the CO-tip for all the AFM simulations \cite{Peng2018} (quadrupole charge of -0.05$\times$0.71$^2$ e$\times$\AA$^2$).

\begin{acknowledgement}

The authors acknowledge insightful discussion with A. Schwarz regarding the interpretation of these data.  This work was financially supported by the Czech Science Foundation (GACR) under Grants No. 15-19672S, No. 17-24210Y, the Purkyne Fellowship and Praemium Academiae program of the Academy of Science of the Czech Republic, and the SPP 1928 ``COORNETS" of the German Science Foundation.

\end{acknowledgement}

\section*{Author Contributions}
P.J, M.S, J.H and A.C devised the experiments.  J.H and A.C made the samples and did the measurements.  B.T and M.M aided the UHV experiments.  G.Z synthesized the molecules.  P.M and P.J did the theoretical calculations. J.H, A.C, and P.J analyzed the data and composed the manuscript.

%\bibliography{bib-file}

\providecommand*\mcitethebibliography{\thebibliography}
\csname @ifundefined\endcsname{endmcitethebibliography}
  {\let\endmcitethebibliography\endthebibliography}{}

\end{document}